Series I

# A note on the Grinberg condition in the cycle spaces

Heping Jiang [0000-0001-5589-808X]

jjhpjhp@gmail.com

## Abstract

Finding a Hamilton graph from simple connected graphs is an important problem in discrete mathematics and computer science. Grinberg Theorem is a well-known necessary condition for planar Hamilton graphs. It divides a plane into two parts: inside and outside faces. The sum of inside faces in a Hamilton graph is a Hamilton cycle. In this paper, using a basis of the cycle space to represent a graph and by the Inclusion-Exclusion Principle, we derive the equality with respect to the inside faces that can be also obtained from Grinberg Theorem. By further investigating the cycle structure of inside faces, we give a new combinatorial interpretation to Grinberg's condition, which explains why Grinberg Theorem is not sufficient for Hamilton graphs. Our results will improve deriving an efficient condition for Hamilton graphs.

## 1 Introduction

A Hamilton cycle is a cycle that contains every vertex of a graph once. A graph is a Hamilton graph if it contains a Hamilton cycle. To determine whether or not a graph is a Hamilton graph is an important subject in discrete mathematics and computer science. Despite of the extensive studies on the problem of Hamilton graphs, as of the present time no good characterization [1, 2] has been found to be an effective condition [3].

In 1968, considering a Hamilton cycle as a Jordan curve on a plane, Emanuel Grinberg characterized the relations of faces in inside and outside the curve and obtained a necessary condition for planar Hamilton graphs [4], which is well-known as follows,

**Grinberg Theorem** *Let G be a planar graph of order* $|V|$ *with a Hamilton cycle C. Then*

$$\sum_{i=3}^{|V|}(i-2)(f'_i - f''_i) = 0, \tag{1.1}$$

*where* $f'_i$ *and* $f''_i$ *are the numbers of faces of degree i contained in inside C and outside C, respectively.*

The equation (1.1) is called the Grinberg condition, the Grinberg formula, or the Grinberg criterion, and is usually used to prove that a graph is non-Hamiltonian.

The follow-up studies on the Grinberg condition almost concentrated on utilizing modular arithmetic to investigate the Hamiltoncity of graphs [5-7]. Nevertheless, note that the given graph G in Grinberg's paper is a planar graph comprised of elementary cycles, which are partitioned into inside faces and outside faces [4]. Then, we can derive an equation associated with inside faces as following the corollary.

**Corollary 1.1** *Let G be a planar graph of order* $|V|$ *with a Hamilton cycle C,* $f'_i$ *the numbers of faces of degree i contained in inside C. then*

$$\sum_{i=3}^{|V|}(if'_i - 2f'_i) = |V| - 2. \tag{1.2}$$

In this paper, we abbreviate the equation (1.2) by the equation and call it the Grinberg condition, unless otherwise indicated. In an algebraic point of view, the corollary implies that the equation has solutions if *G* is Hamiltonian. In other words, in a bounded region on a plane, a Hamilton cycle can be constructed by the faces satisfied the solution, or a Hamilton cycle is the sum of these faces.

Accordingly, such the expression raises a question: Is there a face (or a cycle) structure for a graph that we can use to represent it so that we can convert the structure into the solution faces?

Note that the cycle bases in a cycle space are a compact description of the cycles of a graph and play an important role in studying cycles in graphs [8]. In the beginning of studies in this field, Mac Lane related a linear independent cycle set to a graph in a plane and considered the contraction of a complete set[1] in a graph generated by removing an edge [9]. In 1970's O'Neil further observed the edge removing related to the vector from the basis in a graph [10]. Motivated by their works, we provide a method to obtain the cycles to satisfy the solution from a basis of a graph by removing some cycles (the outside faces). Therefore, if taking inside faces as a solution for the equation, we then have an alternative description of Corollary 1.1.,

**Lemma 1.1** *The equation of a Hamilton graph has solutions.*

According to this lemma, there exist some cycles, in a basis of the cycle space of a Hamilton graph, to build a planar subgraph such that the sum of these cycles is a Hamilton cycle. This indicates that there is no planar limitation for the Grinberg condition to identify the Hamiltoncity of a graph, while in fact every Hamilton graph must have such a planar subgraph. However, the Grinberg condition still remains necessary.

In this paper, we apply the Inclusion-Exclusion Principle to investigate the relations of cycles that construct a Hamilton cycle in a basis, and derive the same equation in Corollary 1.1. Additionally, we scrutinize the combinatorial structure of every two joint cycles, find and characterize that there have two types of combining cycles only for a graph having the solutions. One is that the sum of combining cycles is a Hamilton cycle, while the other is not. We show the new characterization of the Grinberg condition as follows.

**Lemma 1.2** *There only have two types of cycle sets satisfying the solution of a graph G,*

*(1) a 2-common(v, e) combinatorial cycle set, or*

*(2) a 2- common(v,0) combinatorial cycle set.*

---

[1] It implies a basis. See [9].

In the lemma, the type (1) cycle set gives a new combinatorial interpretation to the Grinberg condition, and the type (2) cycle set gives an explanation why the Grinberg condition is not sufficient for Hamilton graphs. Our results will improve deriving an efficient condition for Hamilton graphs.

We will give the definitions and terminologies in section 2, and the proof of Lemma 1.2 in section 3.

## 2 Preliminaries

Graphs considered in this paper are finite, undirected, and simple connected graphs. A graph $G = (V, E)$ consists of $n$ vertices, $V = \{v_1, v_2, \ldots, v_n\}$, and $m$ edges, disjoint from $V$ ($V \cap E = \emptyset$), $E = \{e_1, e_2, \ldots, e_m\}$, together with an incident function (a map) that associates with each edge of $G$ an unordered pair of (necessarily distinct in our paper, and no loops) vertices of $G$.

A subgraph $G' = (V', E')$ of $G$ is a graph with $V' \subseteq V$ and $E' \subseteq E$. A path is a subgraph $P = (V, E)$ of the form $V = \{x_1, x_2, \ldots, x_k\}$ and $E = \{x_1x_2, x_2x_3, \ldots, x_{k-1}x_k\}$, where $x_i$ are all distinct. A graph $G$ is called connected if any two of its vertices are linked by a path in $G$. We define a cycle to be the subgraph $C := P + x_k x_1$ for $k \geq 3$. A cycle is called elementary if no other cycle is properly contained in it. Unless otherwise stated all cycles are elementary in this paper. The size of a cycle is its number of edges and the order of a cycle its number of vertices. A cycle is called a Hamilton cycle that contains every vertex of a graph, and a graph is a Hamilton graph if it contains a Hamilton cycle; sometimes we use Hamiltonian for Hamilton cycles or Hamilton graphs in abbreviation. Two cycles is called joint if they share at least a vertex, sometimes called a common vertex.

For the union of disjoint graphs $A$ and $B$, we define $A \cup B$ with $V(A) \cup V(B)$ and $E(A) \cup E(B)$; the intersection $A \cap B$ is defined analogously. Particularly, when $A$ and $B$ are disjoint, their intersection is the null graph and the cardinality is zero. We denote by $A+B$ =

$(A \cup B) \setminus (A \cap B)$ the sum of subgraphs $A$ and $B$ in a graph.

The edge space of a graph $G$, denoted by $\mathcal{E}(G)$, is a vector space of $m-n+1$ dimension over $GF(2)$ with respect to the operation of sum. The cycle space of $G$, denoted by $CS(G)$, is a subspace of $\mathcal{E}(G)$ generated by edge set of cycles. A cycle basis of $G$ (or a basis in abbreviation), denoted by $\mathcal{B}$(G), is a minimal set of cycles, writes as $\mathcal{B}(G) = \{C_3, C_4, \ldots, C_{|V|}\}$ where the subscripts of $C$ are the cycle size, such that any cycle in a graph $G$ can be represented by a sum of the cycles in the basis $\mathcal{B}$(G). Then $|C_3| \cup |C_4| \cup \ldots \cup |C_{|V|}|$ generates a graph $G$. Let $|C_3|$, $|C_4|$, …, $|C_{|V|}|$ denote the numbers of $C_3$, $C_4$, …, $C_{|V|}$, respectively. Then $|\mathcal{B}(G)| = |C_3|+|C_4|+ \ldots + |C_{|V|}|$. Unless stated otherwise, cycles are referred to as basic cycles in this paper.

We call the equality (1.2) the equation of $G$. A solution of the equation of $G$ is a collection of cycles in a basis of cycle space of $G$ whose union is a Hamilton cycle. A graph is called solvable if it has a solution. For a solvable graph, we can obtain a solution by removing redundant cycles from a basis of cycle space of the graph.

We use the order $i$ to label a cycle $C_i$, where $3 \leq i \leq |V|$, then the expression $|C_3| \cup |C_4| \cup \ldots \cup |C_{|V|}|$ can be represented by $|V_3| \cup |V_4| \cup \ldots \cup |V_{|V|}|$. This means a graph can be represented by a collection of vertex sets. In our paper, such representation means we assume that the vertices of a graph are divided by a set of cycle-like elements. Under the assumption, we apply the Inclusion-Exclusion principle to observe the structure of combining cycles in a cycle basis $\mathcal{B}(G)$ of a Hamilton graph. The principle of inclusion-exclusion is a method to calculate the cardinality of the union of the sets, *i.e.*, $|V_3| \cup |V_4| \cup \ldots \cup |V_{|V|}|$. We have a general form as below,

**The Inclusion-Exclusion Principle** *Let $A_1$, $A_2$, …, $A_n$ be finite sets, then*

$$\left|\bigcup_{i=1}^{n} A_i\right| = \sum_{i=1}^{n}|A_i| - \sum_{1\leq i\leq j\leq n}\left|A_i \cap A_j\right|$$

$$+ \sum_{1\leq i\leq j\leq k\leq n}\left|A_i \cap A_j \cap A_k\right| - \cdots + (-1)^{n-1}|A_1 \cap \ldots \cap A_n|.$$

(2.1)

In observing the cycles combination of a solution with the inclusion-Exclusion Principle, we have the cardinality of intersection of disjoint cycles is zero, and the cardinality of every pair of the joint cycles is 2 either or 1. In the case of the cardinality is 2, every pair of joint cycles, *e.g.*, cycle $A$ and $B$, satisfies $|V(A)|\cap|V(B)|=2$ and $|E(A)|\cap|E(B)|=2$, denoted by a 2-common ($v$, $e$) combination; and satisfies $|V(A)|\cap|V(B)|=2$ and $|E(A)|\cap|E(B)|=0$, denoted by a 2-common ($v$, $0$) combination. A cycle set $S$ is called a 2-common ($v$, $e$) / ($v$, $0$) combinatorial cycle set if every pair of joint cycles in $S$ is a 2-common ($v$, $e$) / ($v$, $0$) combination. In the case of the cardinality is 1, it means there exists a vertex that two or more cycles shared due to their permutation in the cycle combination. However, by the change of a basis, we can convert the case to that of the cardinality is 2. Hence, we can use $|V_i\cap V_j|=2$ to represent the cardinality of the intersection of every pair of joint cycles in a solution.

For notions and terminologies not defined in this paper, please refer to [2, 11, 12, 13].

## 3 The proof of Lemma 1.2

Our proof has two parts. Part I is to prove the existence of a solution under the condition of $|V(C_i)\cap V(C_j)|=2$, and Part II is to show there have uniquely two types of combining cycles that satisfy $|V(C_i)\cap V(C_j)|=2$.

Part I

To prove the first statement, we need to show the equation can be derived from a basis of a Hamilton graph. Let $G$ be a Hamilton graph. Note that the expression $|C_3|\cup|C_4|\cup\ldots\cup|C_{|V|}|$ can be represented by $|V_3|\cup|V_4|\cup\ \ldots\cup|V_{|V|}|$. Then we have $|V_3|\cup|V_4|\cup\ \ldots\cup|V_{|V|}|\ =|V|$, and by the inclusion-exclusion principle, we have

$$|V| = \sum_{i=3}^{|\mathrm{V}|}|\mathrm{V}_i| - \sum_{3\le i<j\le k}^{|\mathrm{V}|}|V_i \cap V_j|$$

$$+\sum_{3\le i<j<k\le|\mathrm{V}|}^{|V|}|V_i \cap V_j \cap V_k| - \cdots + (-1)^{|V|-1}|V_3 \cap V_4 \cap V_5 \cdots \cap V_{|V|}|. \tag{3.1}$$

Since the cardinality of every intersection of disjoint cycles is zero, then we rewrite the equality (3.1) as below,

$$|V| = \sum_{i=3}^{|V|}|V_i| - \sum_{3\le i<j\le|\mathrm{V}|}^{|V|}|V_i \cap V_j|. \tag{3.2}$$

Note that $|V_i \cap V_j|=2$ and the number of pair of joint cycles in the solution is $|B(G)|-1$. Then, $\sum_{3\le i<j\le|\mathrm{V}|}^{|V|}|V_i \cap V_j| = 2(|B|-1)$. Using $|C_3| + |C_4| + \cdots + |C_{|\mathrm{V}|}|$ to replace $|B(G)|$, we obtain,

$$\sum_{3\le i<j\le|V|}^{|V|}|V_i \cap V_j|$$

$$= 2\left(|C_3| + |C_4| + \cdots + |C_{|\mathrm{V}|}| - 1\right)$$

$$= 2\left(\sum_{i=3}^{|V|}|C_i| - 1\right). \tag{3.3}$$

In addition, $\sum_{i=3}^{|V|}|V_i|$ is the sum of vertex subsets in B(G),

$$\sum_{i=3}^{|V|}|V_i| = |V_3| + |V_4| + \cdots + |V_{|V|}|. \tag{3.4}$$

Using $i|\mathrm{C}_i|$ to replace $|V_i|$, we have $|V_3| = 3|C_3|$, $|\mathrm{V}_4| = 4|C_4|$, $\cdots$, $|V_{|\mathrm{V}|}| = |\mathrm{V}| \cdot |C_{|V|}|$. Therefore, the equality (3.4) can be rewritten as follows,

$$\sum_{i=3}^{|V|}|V_i| = 3|C_3| + 4|C_4| + \cdots + |V||C_{|V|}| = \sum_{i=3}^{|V|} i|C_i|. \tag{3.5}$$

Using the equality (3.3) and (3.5) to substitute the correspondent items in the equality (3.2), we derive

$$|V| = \sum_{i=3}^{|V|} i|C_i| - 2\left(\sum_{i=3}^{|V|}|C_i| - 1\right). \quad (3.6)$$

Rearranging the equality (3.6) gives:

$$\sum_{i=3}^{|V|}(i|C_i| - 2|C_i|) = |V| - 2. \quad (3.7)$$

If using the number of inside faces $f'_i$ instead of $|C_i|$, we then obtain

$$\sum_{i=3}^{|V|}(if'_i - 2f'_i) = |V| - 2. \quad (3.8)$$

The equality (3.8) is the same as the equation (1.2). Then, we conclude that the existence of a solution holds under the condition of $|V(C_i) \cap V(C_j)| = 2$.

Part II

In the proof of Part I, we suppose *G* is a Hamilton graph, therefore, the cycle set of the solution is a 2-common (*v*, *e*) combination, which apparently satisfies the condition of $|V(C_i) \cap V(C_j)| = 2$. However, besides $|E(C_i) \cap E(C_j)| = 2$, there exists another the edges combination, $|E(C_i) \cap E(C_j)| = 0$, satisfies the solution of *G* under the condition of $|V(C_i) \cap V(C_j)| = 2$. By the definition, it is a 2-common (*v*, *0*) combination. Obviously, under the condition of $|V(C_i) \cap V(C_j)| = 2$, there only have two cases for $E(C_i) \cap E(C_j)$. Hence, there have only two types of cycle sets satisfy the solutions of a graph, the 2-common (*v*, *e*) combination and the 2 -common (*v*, *0*) combination.

We complete the proof. ■

## Acknowledgements

The author is immensely grateful to the anonymous referees for their useful comments on an earlier version of the manuscript.